# Self-Hybridized Exciton-Polariton Photodetectors From Layered Metal-Organic Chalcogenolates

Bongjun Choi[1], Adam D. Alfieri[1], Wangleong Chen[2], Deep Jariwala[1,2*]

[1]Department of Electrical and Systems Engineering, University of Pennsylvania, Philadelphia, Pennsylvania 19104, United States

[2]Department of Materials Science and Engineering, University of Pennsylvania, Philadelphia, Pennsylvania 19104, United States

[*] Corresponding authors: dmj@seas.upenn.edu

## Abstract

Exciton-polaritons (EPs) arising from strong light-matter coupling offer new pathways for controlling optoelectronic properties. While typically requiring closed optical cavities for strong coupling, we demonstrate that 2D metal-organic chalcogenolates (MOCs), mithrene (AgSePh), with a high refractive index (~2.5) and strong excitons enable self-hybridized polaritons photodetectors (PDs) without top mirrors, simplifying device architecture. Through thickness-tuned multimode polariton engineering, we achieve photodetection of sub-bandgap photons via lower polariton states, validated through reflectance, photoluminescence (PL), and photocurrent spectroscopy with quantitative theoretical agreement. Trap-assisted two-photon absorption enables sustained strong coupling even under sub-bandgap excitation. The polariton dispersion yields ultrafast group velocities (~65 μm/ps), extending exciton diffusion lengths from hundreds of nanometers to several micrometers. Strong-coupling devices demonstrate a 2.38-fold enhancement in photo-to-dark current ratio compared to weak-coupling counterparts, establishing a practical route to polariton-enhanced photodetection and light harvesting.

## Introduction

Extending the spectral response of semiconductor photodetectors (PDs) beyond the material bandgap while simultaneously enhancing carrier transport represents a fundamental challenge in optoelectronics. Conventional approaches rely on chemical doping[1-4], heterostructure engineering[5-7], or quantum confinement effects[8-11], each introducing complexity and often degrading other device parameters. Strong light-matter coupling through exciton-polaritons (EPs)—hybrid quasiparticles that are half-light, half-matter—offers an alternative paradigm for manipulating optical and electronic properties without chemical modification[12-15]. When the energy exchange rate between excitons and photons exceeds their respective dissipation rates, the system enters the strong coupling regime, creating upper and lower exciton-polaritons (UEPs and LEPs) states separated by the Rabi splitting energy[13]. These hybrid states fundamentally reshape the material's optical dispersion, enabling new functionalities not possible in weakly coupled systems[16-24].

EPs have demonstrated remarkable properties, including Bose-Einstein condensation[13,25], narrow band emission from hybrid states[23,26,27], ultralow threshold lasing[22,24,28], and ballistic transport exceeding typical exciton diffusion limits[20,29]. However, their implementation typically requires complex closed optical cavities—either distributed Bragg reflectors (DBR) or metallic mirrors—that introduce significant optical losses and fabrication complexity. Recent work has shown that materials with sufficiently high refractive indices can support "self-hybridized" polaritons through internal constructive interference, eliminating the need for top mirrors[26,27,30,31]. This self-cavity effect occurs when the material thickness supports Fabry-Pérot modes that couple with excitonic resonances, dramatically simplifying device architecture while maintaining strong coupling benefits[31].

Despite extensive research on polariton physics and light emission applications, the integration of EPs into practical photodetection remains largely unexplored. Previous demonstrations of polariton-enhanced photodetection have been limited to closed-cavity organic systems[32,33] or focused solely on theoretical predictions[34,35]. Critical questions remain: Can self-hybridized polaritons enable sub-bandgap photodetection? How does strong coupling affect carrier transport in actual device geometries? What mechanisms sustain polariton formation under sub-bandgap excitation where exciton populations are typically insufficient?

Two-dimensional (2D) materials provide an ideal platform for exploring these questions due to their strong excitonic resonances and tunable optical properties[3,11]. Among emerging 2D semiconductors, metal-organic chalcogenolates (MOCs) represent a unique class combining the processability of organic materials with the electronic properties of inorganic semiconductors. Mithrene (AgSePh), a prototypical MOC, features alternating inorganic Ag-Se layers and organic phenyl groups forming natural multi-quantum-well structures[36-39]. This architecture yields exceptional properties: a wide direct optical bandgap (~2.65 eV) independent of thickness, large exciton binding energy (~0.4 eV), high oscillator strength, and crucially, a high refractive index (~2.5) sufficient for self-cavity formation[27,37,40-42]. Unlike transition metal dichalcogenides (TMDCs) that undergo indirect-to-direct bandgap transitions with thickness[43], mithrene maintains its direct gap across all thicknesses, providing unprecedented design flexibility for polariton engineering[27,44].

In this work, we demonstrate self-hybridized multimode EPs in mithrene PDs, achieving sub-bandgap photon detection without top mirror structures. We reveal that thickness-dependent multimode polariton states, confirmed through complementary spectroscopic techniques, enable photocurrent generation from photons > ~130 nm (~0.55 eV) below the material optical bandgap. Critically, we discover that trap-assisted two-photon absorption sustains strong coupling even under sub-bandgap excitation, addressing a fundamental challenge in polariton photodetection. The polariton dispersion engineering yields group velocities approaching ~65 μm/ps, extending effective exciton diffusion from hundreds of nanometers[45,46] to several micrometers. These enhanced transport properties manifest as a ~2.38-fold improvement in photo-to-dark current ratio compared to weakly coupled devices. Our findings establish a practical route to polariton-enhanced photodetection, demonstrating that strong light-matter coupling can simultaneously extend spectral response and improve carrier collection in simplified device architectures suitable for scalable optoelectronic applications.

## Results and Discussion

### Multimode Self-Hybridized EPs in Mithrene

We begin by investigating the tuned optical properties of mithrene in the strong coupling regime. We exfoliate single-crystalline mithrene flakes and dry-transfer them onto a reflective substrate such as Ag, Au, and DBR (see Methods) (Fig. 1a). Mithrene flakes can induce a large phase shift due to their large refractive index, supporting the Fabry-Pérot cavity mode if the mithrene thickness is sufficiently thick[27,42]. At a mithrene thickness of around 100 nm, only the fundamental mode is supported in the medium. However, when the mithrene layer becomes thicker (>200 nm), the system supports multiple Fabry-Pérot cavity modes (SI Fig. S1). Consequently, excitons in mithrene can couple to multiple cavity modes, resulting in a series of hybrid states (UEP/LEP$_{1, 2, ... n}$), as illustrated schematically in Fig. 1b. When the energy-exchange rate between the material and the cavity photon is faster than cavity ($\gamma_c$) and material ($\gamma_m$) dissipation rates, $2g > (\gamma_c + \gamma_m) / 2$, the system enters the strong-coupling regime, excitons and photons behave as a unified quasiparticle, as the self-cavity mode hybridizes with the mithrene's excitons[13,42].

We calculate the reflectance spectrum in the mithrene/Au system as a function of wavelength and mithrene thickness using the transfer matrix method (TMM, Fig. 1c). The relatively thick mithrene exhibits multiple anti-crossings from multiple LEPs modes, in alignment with the experimental data (★). The dispersion is fit with the coupled oscillator model to extract the coupling strength ($g$) and decay rates of the cavity photons ($\gamma_c$) and exciton ($\gamma_m$). For all cavity mode orders, indicating that the system is in the strong coupling regime with large Rabi splitting (> 650 meV) for all cavity modes (SI Table S1). Since the cavity modes are set by the material thickness, i.e., the effective cavity length, the presence of multiple LEPs modes effectively alters the system's reflectance/absorbance as a function of mithrene thickness[14]. Because of the wavelength resolution limitations of our charge-coupled device (CCD) detector in the normal-incidence setup, the presence of UEPs below 400 nm remained undetectable. Therefore, we measure angle-resolved reflectance of thick mithrene crystals (310 nm) using a spectroscopic ellipsometer, which provides access to the deep ultraviolet (DUV) regime as a function of the angle of incidence (AOI). This approach allows us to observe multiple UEPs states together with multiple LEPs states. The calculated angle-dependent reflectance reveals a pronounced anti-crossing behavior of thick mithrene under transverse electric (TE) illumination (SI Fig. S2), substantiating the strong coupling regardless of AOI. We experimentally identify two UEPs (●) resonances around 300–400 nm (3.10–4.13 eV) regime (blue lines), higher-order mode (★), and LEPs(■) resonances (Fig. 1d). The resonances' peak positions are well aligned with the simulated reflectance spectrum results (red lines). For oblique incidence, since the cavity mode exhibits a dispersive behavior, the polariton resonance modes depend on the in-plane momentum of light, $E_{LEP,UEP}(k_\parallel) \approx E_{LEP,UEP}(0) + \hbar^2 k_\parallel^2 / 2m_{LEP,UEP}$, where the $k_\parallel$ is the in-plane momentum (related to AOI), $\hbar$ is a reduced Planck constant, and $m_{LEP,UEP}$ is the polariton effective mass[13]. Accordingly, polaritonic peaks blueshift with increasing angle, confirming dispersive cavity mode-exciton hybridization. These multiple polaritons change the absorption spectrum of the system significantly[47]. Therefore, as shown in Fig. 1e, a significant amount of sub-bandgap absorption is observed from mithrene, in agreement with TMM calculation due to the change of optical dispersion, which will be discussed later in more detail.

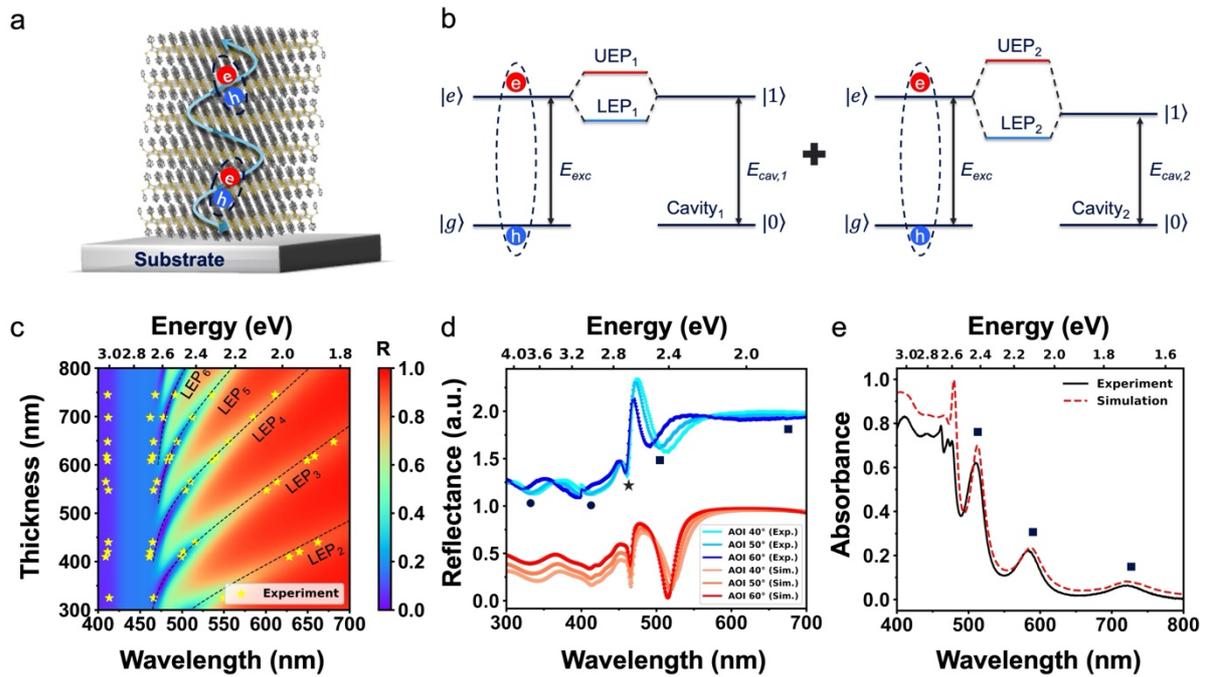

**Figure 1. Self-hybridized multimode EPs in mithrene.** (a) Schematic illustration of self-hybridized EPs on a reflective substrate. Excitons in mithrene hybridize with the multiple Fabry-Pérot cavity modes supported by thick mithrene flakes. (b) Schematic illustration of multiple EPs creation arising from multiple cavity modes in mithrene. Depending on the number of modes supported by mithrene, additional LEPs and UEPs modes can be present. (c) Simulated and experimental reflectance spectra as a function of mithrene thickness on a template-stripped Au substrate, indicating multiple LEPs modes with high agreement between simulation and experimental results. The black dashed lines are the LEPs dispersion calculated by fits to the coupled oscillator model. (d) Simulated (red) and experimental (blue) reflectance spectra as a function of AOI in mithrene (310 nm)/ a template-stripped Au (100 nm)/ Si substrate structure, measured as a function of AOI down to the UV regime (~300 nm), revealing distinct UEPs and LEPs modes with excellent agreement between simulation and experiment. (e) Simulated and experimental absorbance spectra of thick mithrene (~700 nm) on a template-stripped Au substrate reveal significant absorption below the band edge under strong coupling due to the LEPs (■) modes.

## Sub-Bandgap Excitation Induced Strong Coupling

We further examine how the hybridization impacts the optical dispersion in the excitonic regime and sub-bandgap regime. We record photoluminescence excitation (PLE) spectra to investigate the excitonic resonances in the 400−445 nm (2.79−3.10 eV) range under both weak and strong coupling conditions (Fig. 2a). For the weak coupling regime, the thin (~ 30 nm) mithrene crystals are placed on the c-plane sapphire ($Al_2O_3$) substrate since the system cannot support the strong coupling due to the low reflectance from the substrate. Mithrene in the weak coupling regime shows two dominant exciton resonances, potentially from $X_2$ and $X_3$ excitons, which are attributed to the intrinsic exciton resonances from the mithrene[40]. In contrast, under strong coupling, mithrene exhibits a resonance profile that differs from the weak coupling regime, displaying a dominant UEPs mode peak that aligns with the observations in Figs. 1c and d. Moreover, this optical dispersion tuning from self-hybridization can be observed in the emission properties of mithrene since it has a direct bandgap regardless of thickness. Mithrene shows only one intrinsic excitonic photoluminescence (PL) peak around 467 nm (2.65 eV) in weak coupling regime (dashed line) with 405 nm (3.06 eV) continuous wave (CW) laser excitation, on the other hand, when the system enters strong coupling regime the hybrid states change emission properties significantly (Fig. 2b). Multiple reflectance minima at 466, 472, 483, 497, 537, 595, 673 and 787 nm, originating from the LEPs states, give rise to distinct emission peaks as excited electrons relax from the LEPs modes back to the ground state (Fig.2b and SI Fig. S3)[47]. Even these LEPs modes far below the band edges (> 600 nm) produce the emission peak, evident in the 30×-magnified PL spectrum (blue line), substantiating the impact of strong coupling on optical dispersion. Furthermore, these polariton emissions are observed even under the 633 nm (1.96 eV) CW laser excitation, which is a smaller excitation energy than the material's bandgap, indicating that hybridization persists under sub-bandgap excitation (Fig. 2c). In Fig. 2c, sub-bandgap excitation with a 633 nm laser still produces polaritonic emission at the same peak wavelength observed under 405 nm excitation, indicating that sub-bandgap photons can induce the strong coupling in the system. Note that the sharp PL drop below 633 nm originates from the long pass filter. This sub-excitation induced sub-bandgap emission with the identical emission peaks wavelength is observed across the whole flakes with the same spatial pattern under 405 nm laser excitation (Fig. 2d and SI Fig. S4), which enables sub-bandgap absorption under sub-bandgap excitation discussed later.

We further investigate the origin of EPs hybridization under sub-bandgap excitation since EPs hybridization requires a large exciton population along with the cavity mode, which is typically not achieved under sub-bandgap excitation[13]. Using an additional short pass filter, we observe the excitonic regime under sub-bandgap excitation (633 nm). Under sub-bandgap excitation, the PL emission spectrum closely matches that observed under 405 nm laser pumping (Fig. 2e) due to the non-linear optical response in the system, known as upconversion[48]. Power-dependent PL measurement under 633 nm excitation exhibits super-linear behavior, suggesting a nonlinear two-photon process (SI Fig. S5)[49]. Two-photon absorption in mithrene facilitates exciton formation and subsequent hybridization within the system, and we further investigate the underlying mechanism of this two-photon absorption. We reveal that trap sites in mithrene, which produce broad and weak emissions around 725 nm (SI Fig. S6), facilitate the two-photon absorption in the system[41,50]. The broad trap-induced emission is clearly distinct from resonant LEPs-induced emission in the strong coupling regime (SI Fig. S7a). Moreover, the LEPs emission exhibits a thickness dependence because the cavity photon mode is strongly linked to cavity length, which contrasts with the behavior of the trap-induced emission (SI Fig. S7b). Therefore, excitation with sufficient energy to excite the electron into trap states yields the same LEPs resonances as those observed under 405 nm excitation, since trap-assisted two-photon absorption efficiently generates excitons and drives the system into strong coupling. On the other hand, a 785 nm (1.58 eV) laser excitation, which is not enough energy to excite the ground state electrons to trap states, does not generate a sufficiently large exciton population. Therefore, the LEP$_n$ emission (■) around 840 nm observed under 405 and 633 nm excitation is absent under 785 nm excitation since the strong coupling is not achievable (Fig. 2f). Even at higher excitation power of 785 nm excitation, exciton generation in the system remains insufficient, indicating the absence of the LEPs peak around 840 nm (SI Fig. S8). Combining all the observations, we conclude that the trap states play a crucial role in the exciton generation and strong coupling under sub-bandgap excitation in the system. The schematic illustration in SI Fig. S9 shows the sub-bandgap excitation-induced hybridization process. This hybridization induced by sub-bandgap excitation is crucial because it enables tuning of the optical dispersion and enhances the harvesting of sub-bandgap photons.

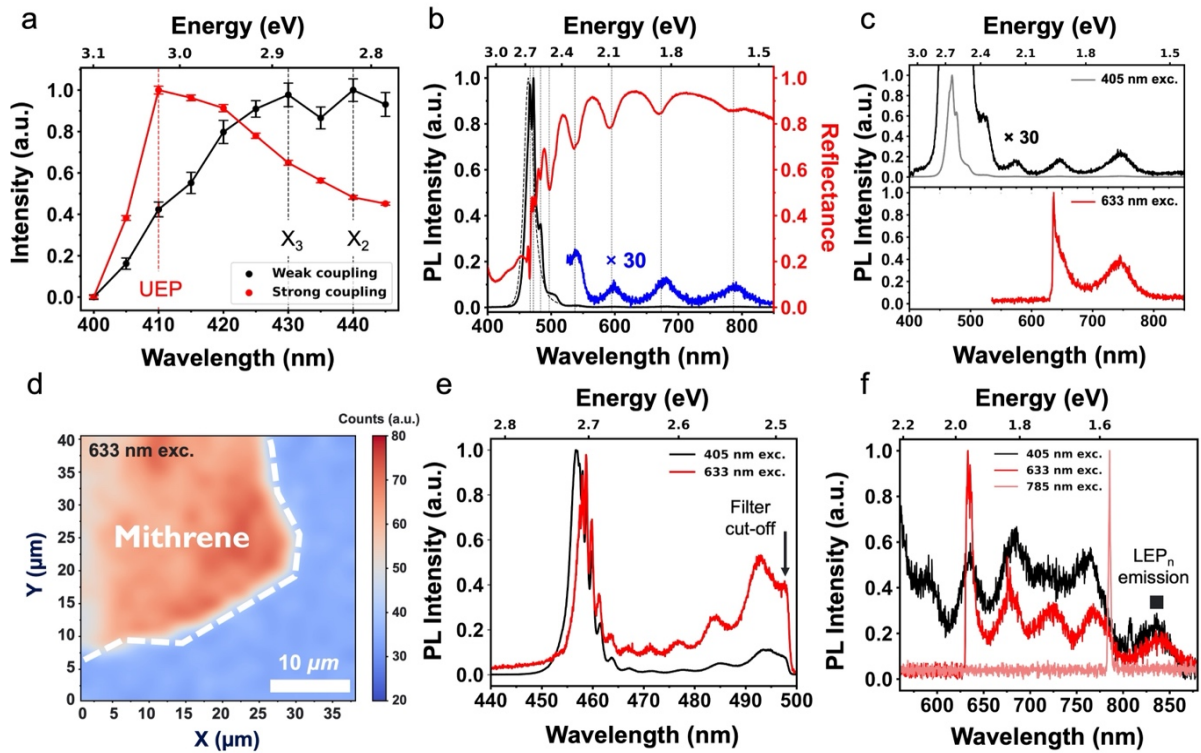

**Figure 2. Optical dispersion tuning in the strong coupling regime and sub-bandgap excitation induced strong coupling.** (a) PLE spectra depending on the coupling regime, indicating optical dispersion tuning in the excitonic regime. Resonance peaks from excitons in the weak coupling regime change significantly in the strong coupling regime. (b) PL and reflectance spectrum in the strong coupling regime clearly show optical dispersion tuning in the excitonic and sub-bandgap regimes. (c) PL spectra depending on the excitation energy, indicating sub-bandgap excitation induced strong coupling. (d) PL map from the LEPs mode under sub-bandgap excitation (633 nm). The LEPs emission peak is observed consistently across all mithrene flakes. (e) PL spectra using a short pass filter with 405 and 633 nm excitation confirm the upconversion via two-photon absorption in mithrene. (f) PL spectra as a function of excitation energy reveal trap-assisted two-photon absorption, and 785 nm excitation does not efficiently produce $LEP_n$ emission (■), unlike 405 and 633 nm excitation.

### Sub-Bandgap Photon Detection in Mithrene Photodetectors

We reveal that strong coupling markedly modifies the optical dispersion by introducing hybridized states and altered absorption characteristics, presenting new possibilities for polaritonic optoelectronic applications. We exploit these effects in PDs to enable the detection of sub-bandgap photons. We design and fabricate (see Methods) two types of two-terminal mithrene PDs, weak vs. strong, by controlling the bottom substrate and thickness of mithrene to study the role of EPs in mithrene PDs[27]. Fig. 3a illustrates the cross-sectional schematic of the mithrene PDs structures depending on the design. Mithrene PDs fabricated on sapphire substrates cannot support strong coupling because they cannot sustain the cavity mode inside mithrene due to the low reflectance from the substrate. In contrast, we employ DBR substrates with sufficiently thick mithrene crystals for strong coupling devices. The fabricated DBR using 9 pairs of $SiO_2/SiN_x$ shows high reflectance around 99% in the excitonic regime of mithrene (SI Fig. S10). Due to its high reflectivity, the mithrene shows clear anti-crossing behavior on the DBR substrate (SI Fig. S11), demonstrating a strong coupling effect. Note that dielectric mirrors are used because metallic substrates cannot be employed due to electrical contact issues, as their conductivity would short the source and drain electrodes. Figure 3b indicates the representative PDs current-voltage ($I_{ds}-V_{ds}$) curves from the weak coupling device under dark and 405 nm laser illumination conditions, indicating a clear photo response of mithrene PDs. Figure 3c shows that photocurrents are overall uniformly distributed across the entire channel regime of the device under 5V bias, owing to the large drift from the electric field.

Next, we examine how strong-coupling affects mithrene PDs in sub-bandgap photon absorption by comparing the weak and strong coupling devices. We first measure the reflectance of each system to verify the LEPs modes induced absorption, and subsequently the photocurrent as a function of incident wavelength using a photocurrent spectroscopy measurement setup under 5V bias. As shown in the top panel of Figure 3d, the weak coupling device does not show the dominant resonance below the bandgap of mithrene in the $1 - R(\lambda)$ spectrum (pseudo-absorption spectrum) with good agreement with the TMM simulation results (black dashed line). The weak coupling device shows dominant absorption around 460 nm from the exciton resonance, which is aligned with prior literature[27,41]. Consequently, the photocurrent below the sub-bandgap is negligible and exhibits only minor tails likely from Urbach absorption (Fig. 3d bottom panel)[51]. In contrast, the strong coupling devices show a distinct reflectance and photocurrent trend compared to the weak coupling device. The $1 - R(\lambda)$ spectrum reveals multiple absorption resonances originating from lower polariton states below the bandgap regime, highly aligned with the theoretical prediction (black dashed line) as shown in Fig. 3e. Consequently, the strongly coupled device exhibits pronounced photocurrent in the sub-bandgap regime arising from the LEPs states (■) induced absorption (red-shaded regime). The spectral shape of the wavelength-resolved photocurrent spectrum of the PDs is largely dictated by the optical dispersion in the system, showing a similar trend regardless of applied bias (SI Fig. S12). In strongly coupled devices, sub-bandgap photocurrent follows the LEPs resonances, unlike the weak device's Urbach tail, well aligned with the observation in Figs. 1e and 2b. Notably, under sub-bandgap photoexcitation (~ >465 nm, <2.66 eV), strong coupling persists via a two-photon absorption process proven in Fig. 2e, enabling sub-bandgap photodetection.

The absorption-peak position can be tuned by adjusting the cavity length, as the system's optical dispersion shifts accordingly[14]. This approach yields detection wavelength-tunable PDs without any chemical modification and can be applied to materials with strong exciton resonances besides the mithrene system[19]. We demonstrate this tunability by showing that devices with varying mithrene thicknesses exhibit distinct absorption peak positions in the sub-bandgap regime. (SI Fig. S13). Using the TMM, we calculate the absorption contributed solely by the mithrene layer as a function of its thickness to estimate the absorption peak. Figure 3f shows that the peak positions extracted from the current ($I_{ds}$) vs. wavelength curves in multiple devices with different mithrene thickness (SI Fig. S13) closely match the calculated results, indicating that sub-bandgap absorption from LEPs states generates the photocurrent. This observation shows that sub-bandgap photons can be harvested and detected by inducing strong coupling in the system without any chemical modification of the photoactive layer.

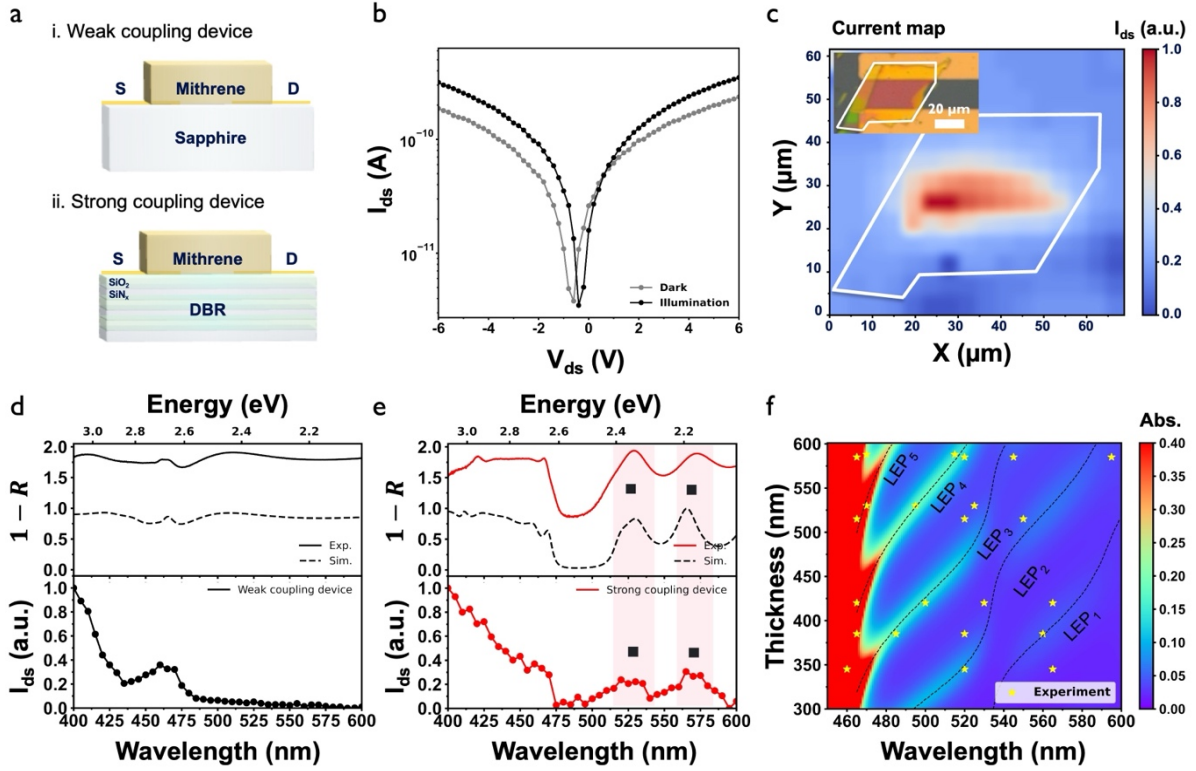

**Figure 3. Sub-bandgap absorption in mithrene PDs.** (a) Schematic illustration of mithrene PDs depending on the (i) weak and (ii) strong coupling configuration. (b) Representative current vs. voltage ($I_{ds}$-$V_{ds}$) curve from a weak coupling device, indicating a clear photo response from mithrene. (c) Photocurrent mapping from a weak-coupling device with 405 nm CW laser illumination. Inset is the optical microscopy (OM) image of representative mithrene PDs used for photocurrent mapping. The experimental (solid line) and simulated (dashed line) $1 - Reflectance(\lambda)$ spectra (top panel), together with the normalized wavelength-resolved $I_{ds}$ photocurrent ($I_{ds}-\lambda$) spectra (bottom panel) for representative (d) weak-coupling and (e) strong-coupling devices, showing the distinct photo response depending on the design, especially below the bandgap regime. (f) Simulated absorbance spectra solely from the mithrene layer using TMM, plotted as a function of mithrene's thickness on the DBR substrate, are overlaid with the experimental $I_{ds}$ peak from the wavelength-resolved $I_{ds}$ spectrum. The close agreement between simulation and experiment indicates the generation of sub-bandgap induced photocurrent. The black dashed lines show the multiple LEPs dispersions in the system.

## Enhanced Photocurrent Generation in the Strong Coupling Regime

The photonic component gives EPs an ultralight effective mass ($\sim10^{-4} \cdot m_e$), high group velocity, and thus long-range transport within the cavity[13]. Their delocalized character also enables enhanced exciton diffusion and energy transfer compared with the Förster mechanism[33,45,52]. Figure 4a presents the dark current and photocurrent of weakly and strongly coupled devices of similar mithrene thickness under a non-resonant 405 nm CW laser pump. For these measurements, we used a 100× objective (numerical aperture (N.A.) = 0.90) to focus the beam to an approximately 3 μm diameter spot, which is much smaller than the 40 μm channel length, meaning that exciton diffusion plays a significant role for photocurrent generation. For the dark current, the current level is similar between weak and strong-coupled devices since this dark current is mostly attributed to free carrier injection[53]. Under illumination, current is dominated by exciton diffusion (given the large exciton binding energy), and the strong-coupling device shows a larger increase than the weak-coupling device due to the enhanced exciton transport discussed later. Note that the absorption from the mithrene layer mostly remains the same for 405 nm illumination, regardless of substrate (SI Fig. S14); therefore, the increase does not originate from an absorption increase. Moreover, this increase is seen across multiple devices and varies with both the coupling regime and the excitation objective's N.A. (Figs. 4b and c). The strongly coupled device exhibits a 2.38-fold higher photo-to-dark current ratio ($I_{illumination}/I_{dark}$) than its weakly coupled counterpart under the same bias conditions, indicating more efficient photocurrent generation. Interestingly, when the excitation N.A. is increased from 0.35 (corresponding to a maximum half-angle of 20.49°, $k_\parallel \sim 4.5\ \mu m^{-1}$) to 0.90 (64.16°, $k_\parallel \sim 14\ \mu m^{-1}$), the photo-to-dark current ratio enhancement grows since a larger excitation cone promotes longer-range EPs transport in the strong coupling regime, consequently, higher photocurrent. Together, these results demonstrate that strong coupling in PDs plays a critical role in maximizing photocurrent response.

To gain more insight into the photocurrent enhancement, we further elucidate the role of strong coupling and incidence angle dependence in photocurrent generation. We calculate the theoretical group velocity ($v_g = \partial \omega / \partial k_\parallel$) of LEP$_1$ mode as a function of in-plane momentum of light ($k_\parallel$), based on the dispersion relation of our systems in SI Fig. S11. The calculated $v_g$ indicates an increasing trend as the in-plane momentum of light increases (Fig. 4d). At the maximum $v_g$ reaches of the $\sim$65 $\mu m/ps$ ($\sim$0.22$c$), where c is the speed of light. By collecting light over a wider cone (higher N.A. $\sim$ 0.90), it simultaneously excites a broader range of in-plane momenta than a smaller N.A. $\sim$ 0.35, which matches the EPs dispersion; this increases the population of polaritonic modes with higher $v_g$ that can propagate across the self-cavity. Moreover, the increase in $v_g$ impacts the propagation length of EPs. We calculate the theoretical EPs' propagation length ($L_{EPs}$) in our systems by combining the group velocity ($v_g$) with the effective polariton lifetime ($\tau_{LEP}$), which itself is set by the photon and exciton Hopfield fractions in Eq. (1)[54]:

$$L_{LEP} \approx v_g \tau_{LEP} \approx v_g \left( f_{cav}/\tau_{cav} + f_{exc}/\tau_{exc} \right)^{-1} \tag{1}$$

where $f_{cav(exc)}$ is the Hopfield coefficient of cavity photon (or exciton) in the system ($f_{cav} + f_{exc} = 1$) extracted using the two-coupled oscillator models as a function of in-plane momentum[13], and $\tau_{cav(exc)}$ is the lifetime of cavity photon (exciton)[54,55] (SI Fig. S15). Figure 4e indicates the LEPs propagation distance as a function of in-plane momentum; at higher $k_\parallel$, LEPs can travel several micrometers. This represents an order-of-magnitude enhancement compared to the typical exciton diffusion length (a few hundred nanometers[45,46]) and underscores the dramatic transport benefit conferred by strong light–matter coupling. Moreover, because the polariton lifetime scales with the photon lifetime, this boost in propagation length correlates strongly with the cavity's quality factor[34]: cavities with higher Q-factor produce longer-lived polaritons and thus enable farther transport. Figure 4f shows the propagation length depending on the Q-factor. By increasing the Q-factor, polariton propagation length could be extended to the order of tens of μm. Open-cavity systems typically support only relatively low Q-factors because they form lossy Fabry–Pérot modes, unlike closed cavities[56]. Building a high-Q system with two mirrors can enhance transport, but this comes at the cost of increased absorption loss from the top mirror. Alternatively, patterning mithrene into a photonic-crystal structure can achieve Q-factors on the order of thousands without a top mirror and sustain transport lengths of several tens of micrometers[57]. This is beyond the scope of this paper but should be explored in future work.

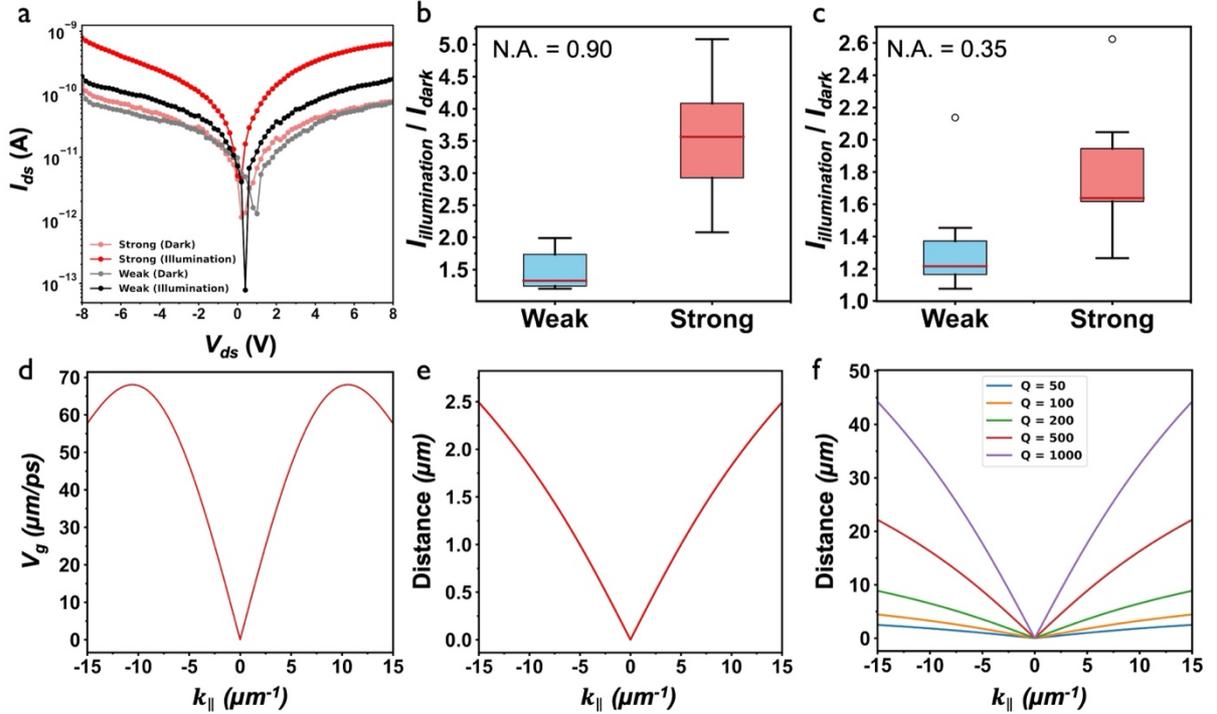

**Figure 4. Enhanced photocurrent facilitated by the EPs transport in mithrene PDs.** (a) Representative current vs. voltage curves for weak and strong coupling regimes under both dark and illuminated conditions. Box plots of photo-to-dark current ($I_{illumination}/I_{dark}$) ratio for weak and strong coupling devices, depending on the excitation objective's N.A. (b) N.A. ~ 0.90 and (c) N.A. ~ 0.35 under 5V bias (circle indicates the outliers) from 9 weakly coupled and 6 strongly coupled devices, indicating the largest enhancement in strong coupling and high N.A. excitation. (d) Calculated theoretical group velocity ($v_g$) of EPs as a function of in-plane momenta ($k_\parallel$) from the dispersion relation in mithrene (300nm) on the DBR structure. (e) Calculated theoretical polariton propagation length as a function of in-plane momenta, Hopfield coefficients are extracted using a two-coupled oscillator model (SI Fig. S15), and the exciton lifetime is adapted from previous literature[42]. (f) Polariton propagation length depending on the cavity Q-factor as a function of in-plane momenta, exhibiting a longer propagation length with a high Q-factor.

## Conclusion

In summary, we have demonstrated self-hybridized multimode EPs in 2D mithrene, enabling practical PDs applications without requiring top mirror structures. The key achievements include: (1) Thickness-tunable multimode polariton states verified through comprehensive optical spectroscopy with quantitative theoretical agreement; (2) Sub-bandgap photodetection extending response > ~130 nm (~0.55 eV) below the material optical bandgap through lower polariton absorption; (3) Enhancement in photo-to-dark current ratio in strong-coupling devices due to the enhanced exciton transport predicted from the polariton dispersion. These findings establish self-hybridized polaritons as a viable approach for extending PDs' spectral response and enhancing carrier collection without chemical modification. The simplified architecture, eliminating top mirrors, reduces optical losses while maintaining strong coupling benefits. Future work should explore voltage-tunable coupling strength for dynamic spectral control and patterning of mithrene crystals into 1D grating or 2D photonic-crystal structures. This work paves the way for a new class of polariton-enhanced optoelectronic devices operating beyond conventional bandgap limitations.

## Methods

### Mithrene synthesis and sample preparation.

We synthesized mithrene crystals via an organic single-phase method incorporating propylamine $(PrNH_2)$[27,38]. The as-grown crystals were transferred onto various substrates (e.g., c-plane sapphire $(Al_2O_3)$, Au, and $SiO_2/SiN_x$ DBR) by mechanical exfoliation and dry transfer using polydimethylsiloxane (PDMS). Flake thickness was then characterized using atomic force microscopy (AFM) and ellipsometry.

### Mithrene photodetectors fabrication.

The source and drain electrodes were patterned using photolithography (SUSS MicroTec MA6) on c-plane $Al_2O_3$ and DBR substrates with channel lengths of 20 or 40 μm. Ti (10 nm)/Au (50 nm) were deposited by electron-beam physical vapor deposition (PVD) (Kurt J. Lesker) and lifted off in KL Remover (Kemlab). DBR substrates were fabricated using plasma-enhanced chemical vapor deposition (PECVD) (Oxford PlasmaLab 100), depositing nine alternating pairs of $SiO_2$ (~78 nm)/$SiN_x$ (~56 nm) on a Si substrate at 350°C. Mithrene flakes were then dry-transferred onto the pre-patterned electrodes, which were then wire-bonded to a custom-designed printed circuit board (PCB) for electrical measurements.

### Optical characterization (PL, PLE, and reflectance).

The PL, PLE, and reflectance spectra were collected using the Horiba LabRam HR Evolution confocal microscope. PL spectra were recorded under CW laser excitation at 405, 633, and 785 nm. A 100 grooves/mm grating was used before detection by the CCD. For the PLE measurement, a supercontinuum laser source (NKT Photonics) was used with a 450 nm long pass filter with 5 nm spectral resolution. The tunable output was fiber-coupled into the Horiba confocal microscope. The reflectance was measured using the white light source (AvaLight-HAL) with a 50x objective (N.A. = 0.35). A silver mirror was used for the reflectance normalization. Angle-resolved reflectance was obtained with the Accurion EP4 system (Park Systems) using ellipsometric contrast microscopy (ECM) mode. All optical measurements are done at room temperature.

### Optical simulation

Theoretical reflectance and absorbance are calculated using the TMM calculation with homemade Python code[58,59].

**Photocurrent measurement.**

Wavelength-resolved photocurrent was obtained with a homemade setup. The collimated light was generated from the Xe arc lamp light source (Sciencetech), and a monochromator (Newport) was used for the wavelength-resolved photocurrent measurement with 5nm spectral resolution. The incident light power density was measured using a power meter (Thorlabs). The collimated light was shining on the sample, and the photocurrent was measured using a Keithley 2450 sourcemeter. For the incidence-angle-dependent photocurrent measurement, the Horiba LabRam HR confocal was used with connection to a sourcemeter (Keithley 2450) with 50x objective (N.A. = 0.35) and 100x objective (N.A. = 0.90) with 5.9 mW of 405 nm laser power. For the photocurrent mapping, motorized XY stages in the Horiba confocal setup were used with step sizes of approximately 4 μm.

**Acknowledgment.**

D. J., B.C. acknowledges support from the Office of Naval Research Young Investigator Award (N00014-23-1-203), Metamaterials program. D.J. and A.A. also acknowledge partial support from National Science Foundation DMR 2429281. A.A. acknowledges partial support for the work from the Vagelos Institute of Energy Science and Technology Graduate Fellowship.

**Data availability**

All data supporting the findings of this study are provided in the main article and the supplementary information. Additional information is available from the corresponding authors upon reasonable request.

# Self-Hybridized Exciton-Polariton Photodetectors From Layered Metal-Organic Chalcogenolates

Bongjun Choi[1], Adam D. Alfieri[1], Wangleong Chen[2], Deep Jariwala[1,2*]

[1]Department of Electrical and Systems Engineering, University of Pennsylvania, Philadelphia, Pennsylvania 19104, United States

[2]Department of Materials Science and Engineering, University of Pennsylvania, Philadelphia, Pennsylvania 19104, United States

[*] Corresponding authors: dmj@seas.upenn.edu

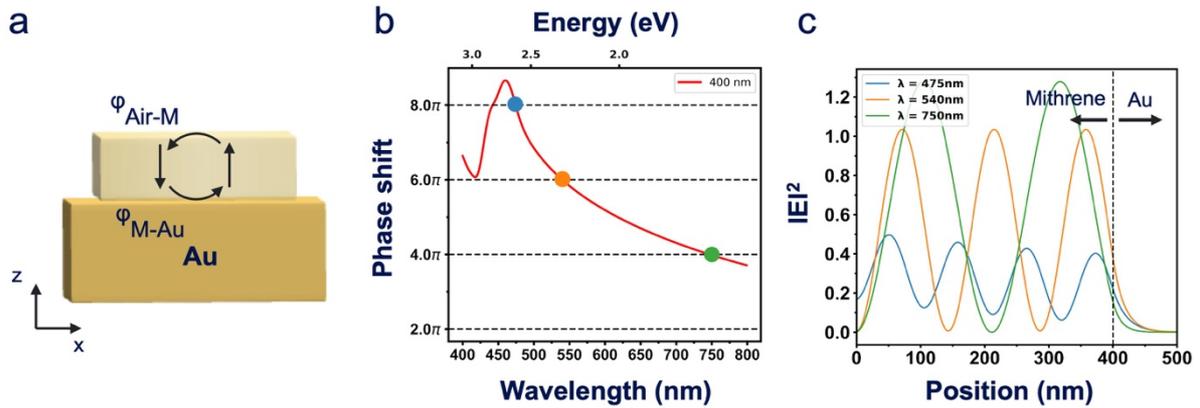

**Figure S1.** Multiple self-cavity modes in mithrene. (a) Schematic illustration of the phase shift in the open cavity system (Mithrene on Au), highlighting the phase changes at the Air-Mithrene and Mithrene-Au interfaces. The phase shift within the mithrene layer is proportional to its refractive index. (b) Calculated total phase shift ($\varphi$) in mithrene (400 nm)/ template-stripped Au structure, indicating the occurrence of a large phase shift due to the large refractive index of mithrene[1,2]. Fabry-Pérot modes can form when the total phase shift becomes an integer multiple of $2\pi$. In the case of the mithrene (400 nm) / template-stripped Au structure, multiple resonant modes are supported, particularly in the low-loss regime, around wavelengths of approximately 475, 540, and 750 nm. (c) Simulated electric field as a function of z-position, showing distinct resonances within the mithrene layer (0-400 nm) that originate from the Fabry-Pérot mode aligned with the results in (b). The electric field decays rapidly in the Au layer (400-500 nm).

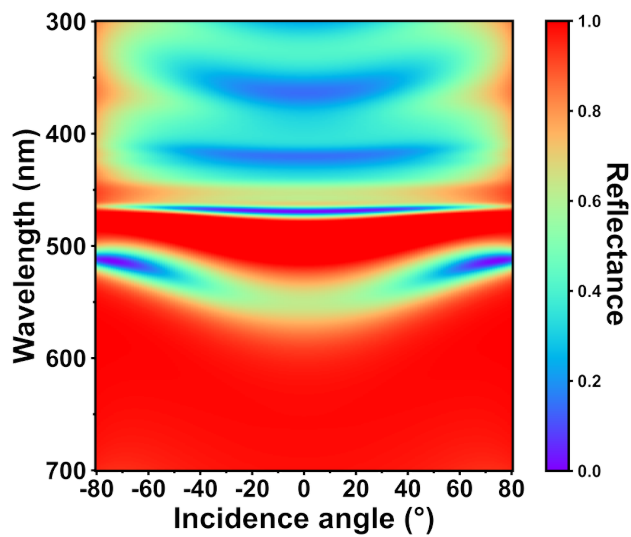

**Figure S2.** Simulated angle-resolved reflectance under transverse electric (TE) illumination in mithrene (310 nm)/template-stripped Au structure, showing clear anticrossing behavior.

| Param. Order | $g$ (meV) | $E_x$ (eV) | $\gamma_m$ (meV) | $\gamma_c$ (meV) | Rabi Splitting | Coupling regime |
|---|---|---|---|---|---|---|
| n = 1 | 352 | 2.79 | 72 | 208 | 690 | Strong |
| n = 2 | 333 | 2.73 | 52 | 142 | 660 | Strong |
| n = 3 | 359 | 2.74 | 53 | 144 | 712 | Strong |

**Table S1.** Multiple EPs modes in mithrene. The parameters are extracted from the two-coupled oscillators model[3].

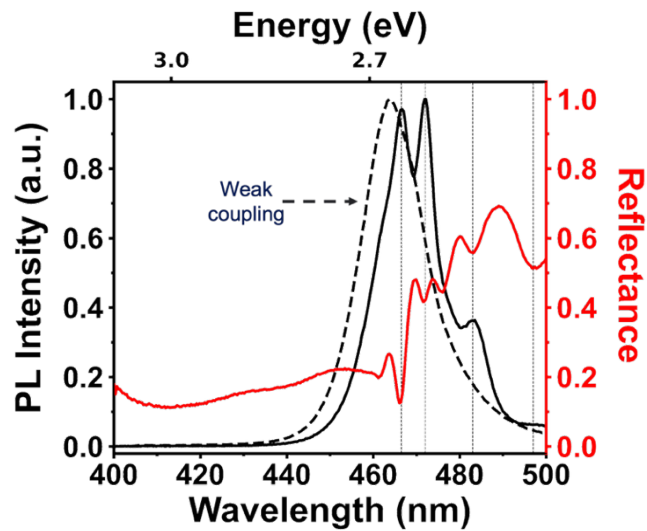

**Figure S3.** Photoluminescence (PL) and reflectance spectra in the strong coupling regime from Figure 2b, focused around the excitonic region, reveal a distinct difference depending on the coupling regime (strong vs. weak).

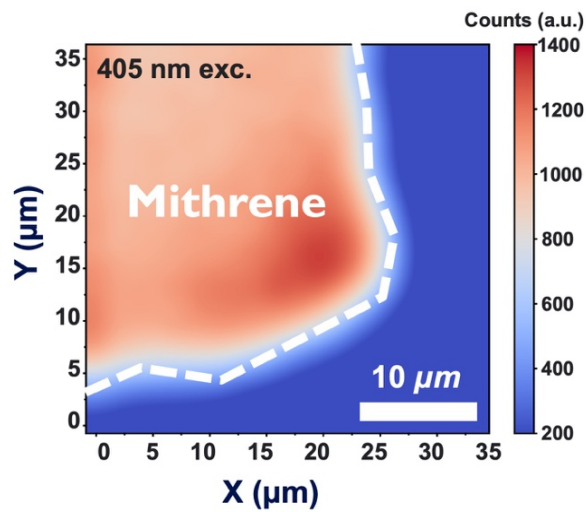

**Figure S4.** PL map obtained from the LEP mode with 405 nm continuous wave (CW) laser excitation. Sub-bandgap excitation induces sub-bandgap emission across the entire mithrene flake.

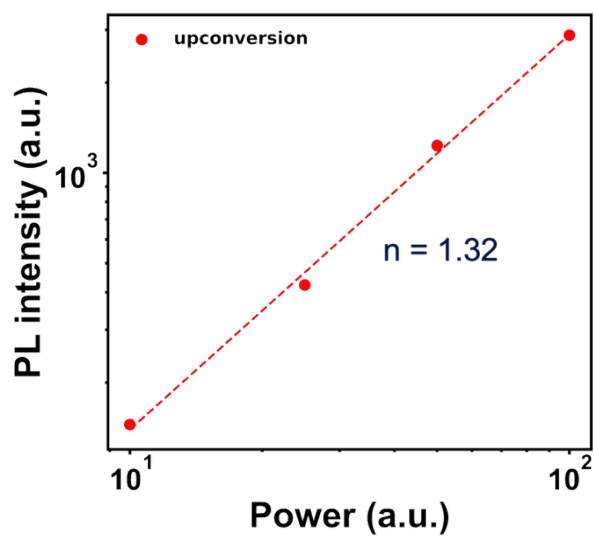

**Figure S5.** Power-dependent PL measurement under 633 nm excitation. The fitted slope (*n*) of the power dependence ($I \propto \mathfrak{a}P^n$) is approximately 1.32, indicating that the emission arises from a two-photon process[4,5].

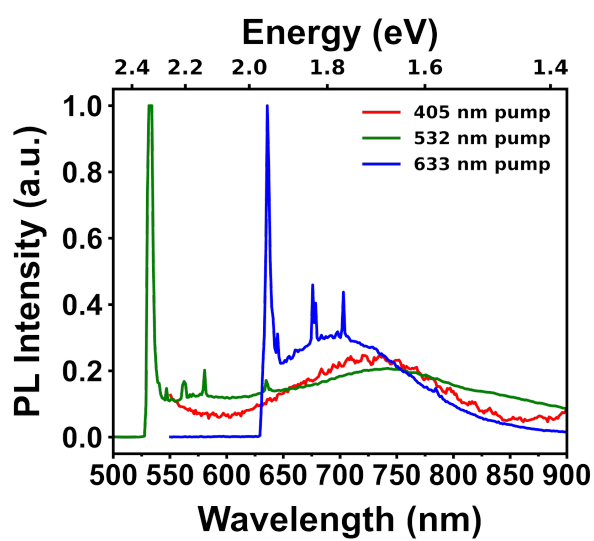

**Figure S6.** Defect-induced emission from mithrene under various excitation wavelengths (405, 532, and 633 nm), showing broad emission around 725 nm. The sharp peaks under 532 and 633 nm excitation originate from Raman scattering in mithrene[6].

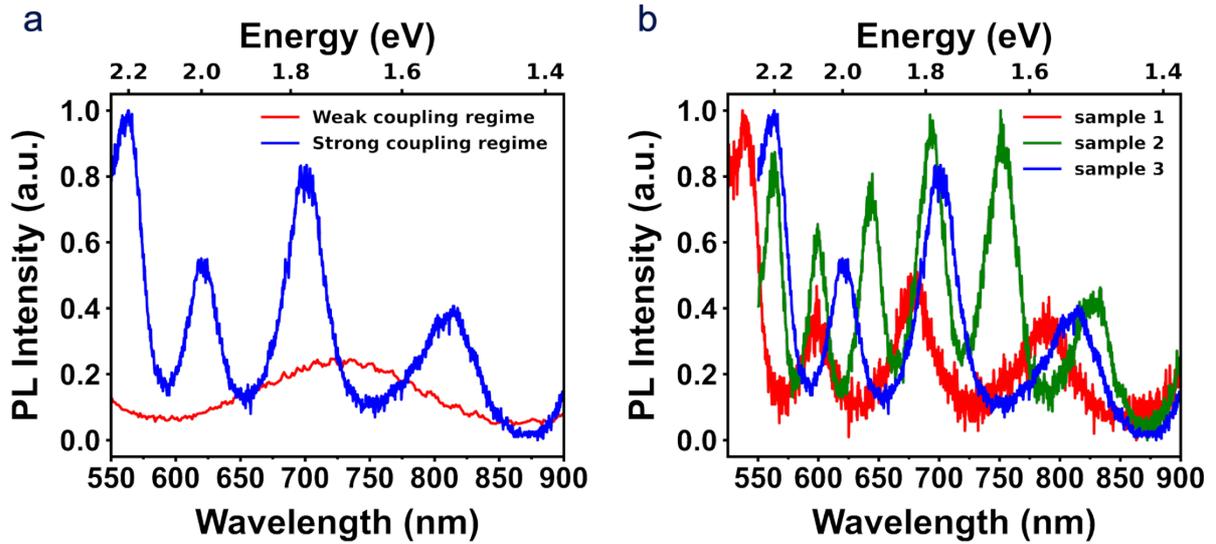

**Figure S7.** (a) PL emission spectra depending on the coupling strength. In the strong coupling regime, mithrene shows multiple resonant peaks due to the LEPs-induced emission; in contrast, the weak coupling regime shows broad emission originating from the trap sites[7]. (b) In the strong coupling regime, the resonant PL peaks vary with mithrene thickness, as mithrene acts as a self-formed cavity. The resulting cavity modes shift with thickness, showing a clear contrast with defect-related emission.

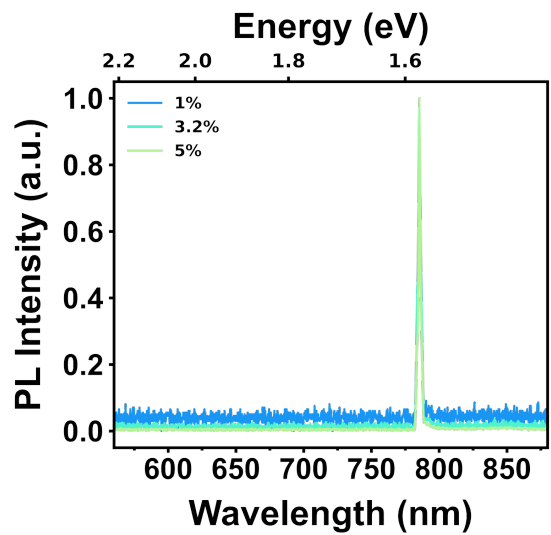

**Figure S8.** PL emission spectra under 785 nm excitation as a function of excitation power. High excitation power does not lead to $LEP_n$ emission, indicating the presence of trap-assisted two-photon absorption in the system.

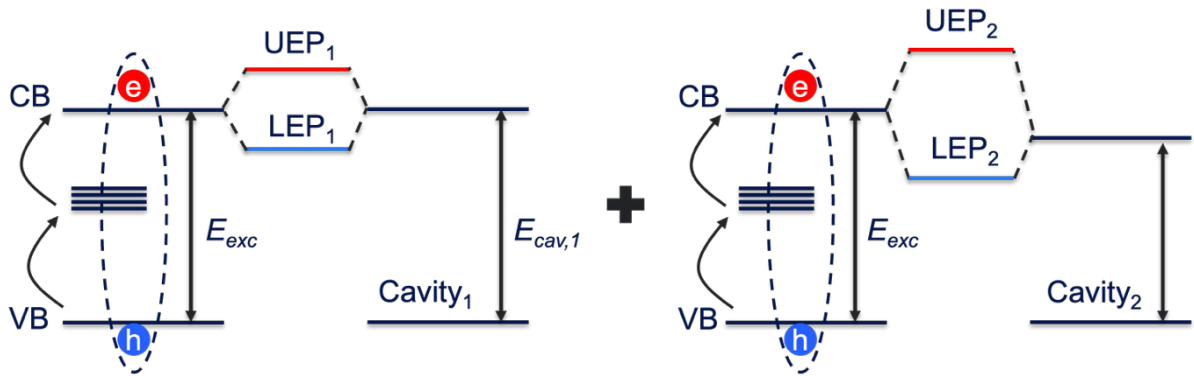

**Figure S9.** Suggested mechanism for sub-bandgap-excitation-induced multiple EPs hybridization. Trap states facilitate the generation of a large population of excitons, and subsequently, hybridization occurs.

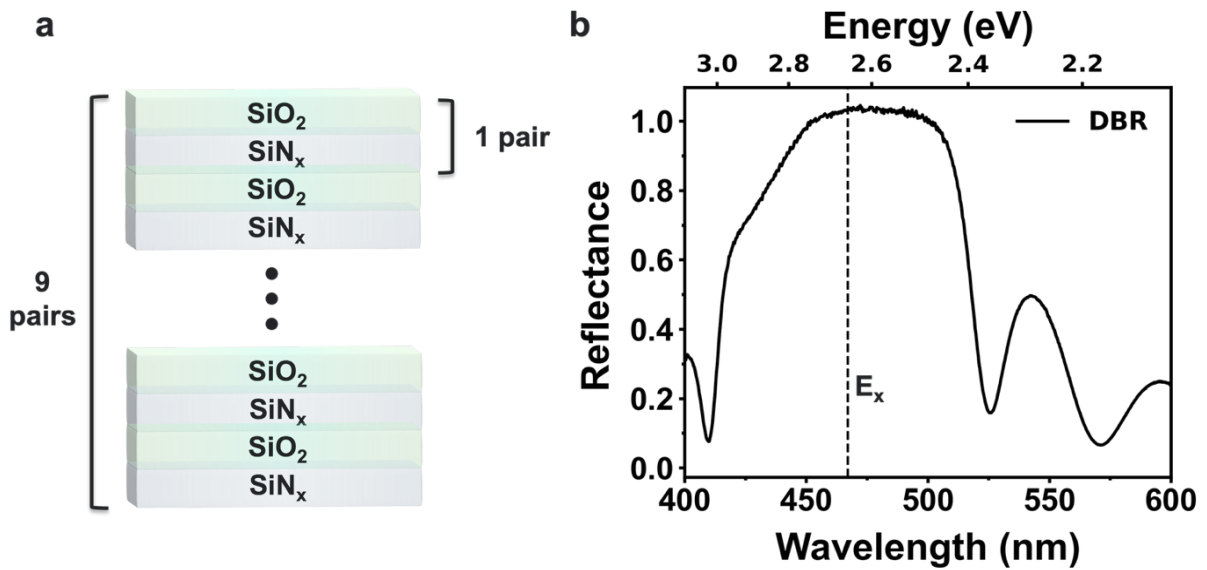

**Figure S10.** (a) Schematic illustration of a distributed Bragg reflector (DBR) made of 9 pairs of $SiO_2/SiN_x$ on a Si substrate. (b) The representative reflectance spectrum of DBR, showing higher reflectance than an Ag substrate in the excitation regime (~450-500 nm).

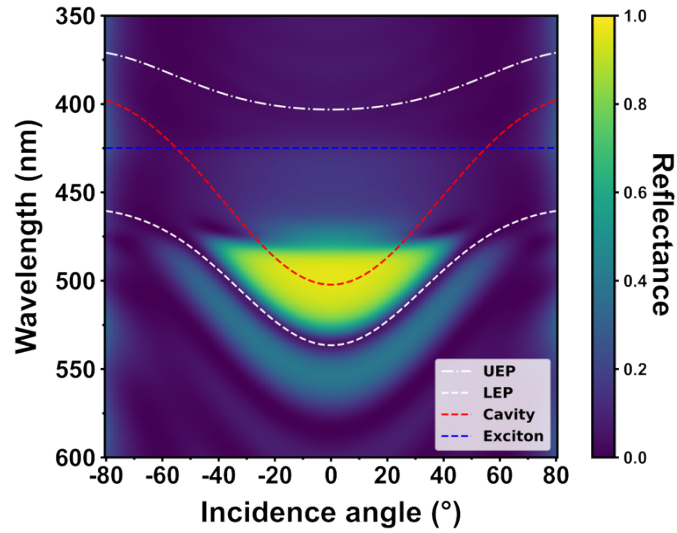

**Figure S11.** Simulated reflectance spectrum as a function of incidence angle in mithrene (300 nm) / DBR structure, indicating clear anticrossing behavior fitted by the two coupled oscillator model.

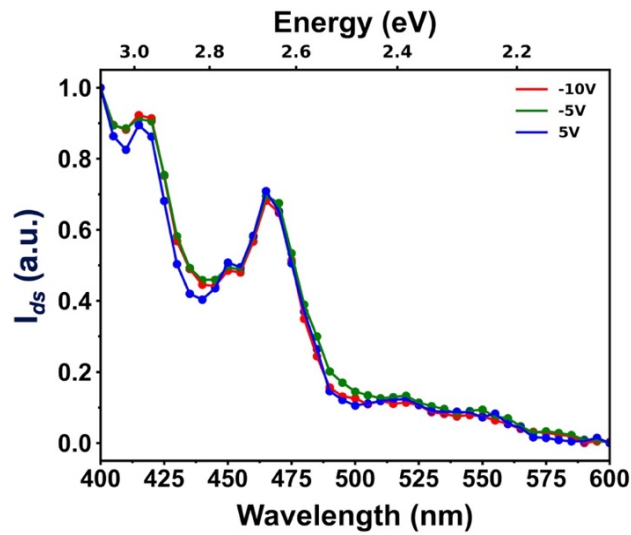

**Figure S12.** Normalized wavelength-resolved $I_{ds}$ photocurrent ($I_{ds}$-$\lambda$) under various bias conditions.

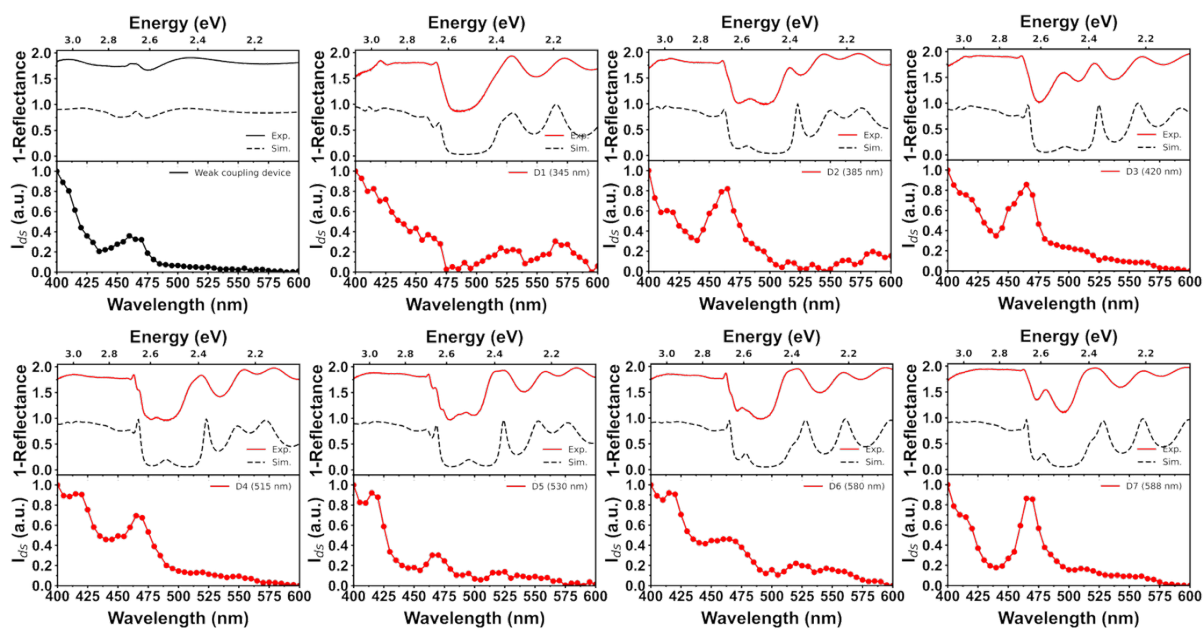

**Figure S13.** The experimental (solid line) and simulated (dashed line) $1 - Reflectance(\lambda)$ spectra (top panel), along with the normalized wavelength-resolved $I_{ds}$ photocurrent ($I_{ds}$-$\lambda$) spectra (bottom panel) from multiple strongly coupled devices.

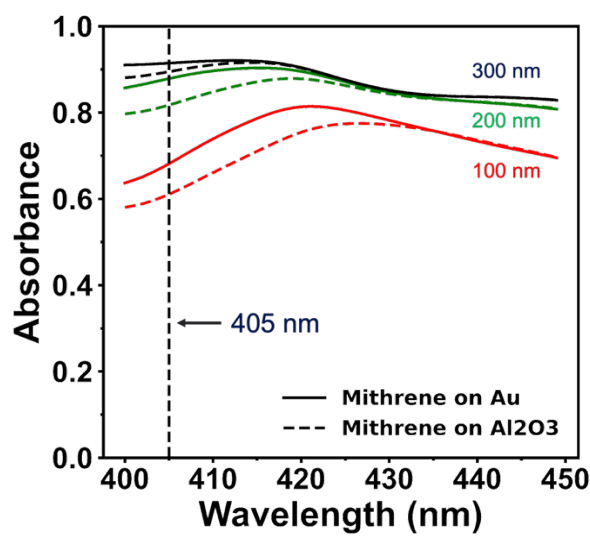

**Figure S14.** Simulated absorbance spectrum arising only from the mithrene layer in multiple mithrene thicknesses (100, 200, and 300 nm), depending on the substrate (Au and Al$_2$O$_3$), indicating similar absorbance under 405 nm laser excitation. As the mithrene thickness increases, the difference becomes less pronounced.

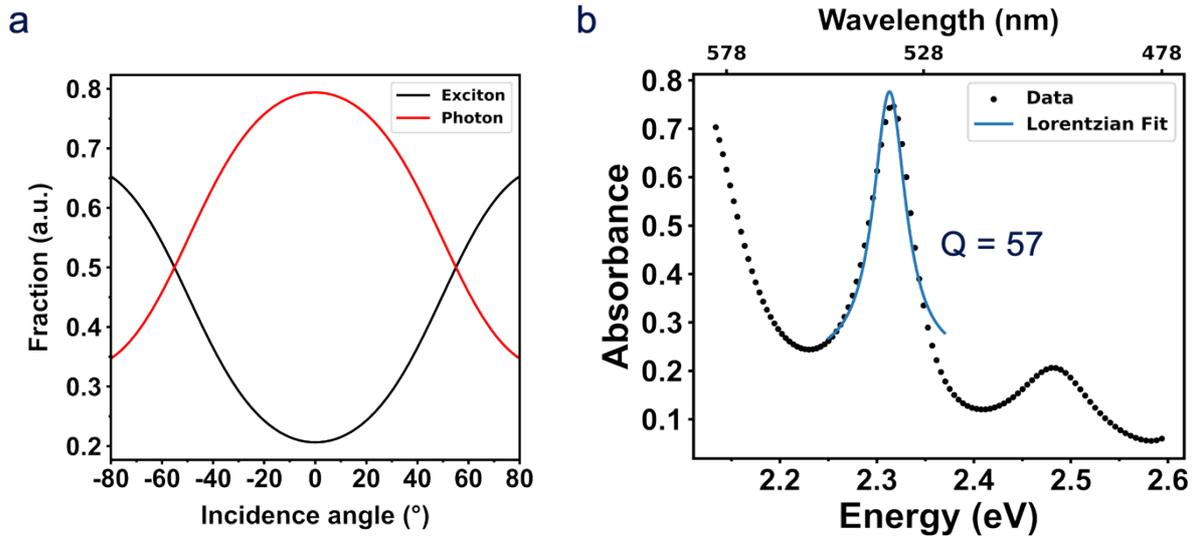

**Figure S15.** (a) Hopfield coefficient as a function of incidence angle, obtained from the dispersion relation in Fig. S11. (b) Simulated absorbance spectrum of a mithrene (300 nm)/ DBR structure, used to estimate the Fabry-Pérot cavity quality factor.